\documentclass[showpacs,amsmath,amssymb,aps,prd,preprint,groupedaddress,superscriptaddress,nofootinbib]{revtex4-2} 
\usepackage[utf8]{inputenc}
\usepackage{graphicx}
\usepackage{tensor}
\usepackage{bm}
\usepackage{color}

\newcommand{\connection}[1]{\tensor{\Gamma}{#1}}
\newcommand{\potential}[1]{\tensor{\Sigma}{#1}}
\newcommand{\pd}[1]{\tensor{\partial}{#1}}
\newcommand{\nablab}{\bm{\nabla}}

\newcommand{\e}[2][]{\tensor{#1{e}}{#2}}

\newcommand{\teta}[2][]{\tensor{#1\vartheta}{#2}}

\newcommand{\bracket}[1]{\left<#1\right>}
\newcommand{\torsion}[1]{\tensor{T}{#1}}

\newcommand{\sconnection}[2][]{\tensor{#1\omega}{#2}}

\newcommand{\energy}[1]{\tensor{t}{#1}}

\newcommand{\K}[1]{\tensor[^3]{K}{#1}}
\newcommand{\tri}[2]{\tensor[^3]{#1}{#2}}
\newcommand{\etime}[2][]{\tensor[#1]{\tau}{#2}}

\newtheorem{theorem}{Theorem}[section]

\begin{document}


\title{The generalization of the ADM gravitational energy-momentum}



\author{J. B. Formiga}
\email[]{jansen@fisica.ufpb.br}
\author{V. R. Gon\c calves}
\affiliation{Departamento de F\' isica, Universidade Federal da Para\' iba, Caixa Postal 5008, 58051-970 Jo\~ ao Pessoa, Pb, Brazil}



\begin{abstract}
In this paper, it is proved that the teleparallel energy-momentum generalizes that of the ADM formalism. In doing so, it is shown that the teleparallel $4$-momentum can be made to coincide with that of the ADM approach whenever the ADM $4$-momentum is applicable. The only assumptions are the time gauge for the teleparallel frame and the well-known restrictions for the coordinate system used in the calculation of the  ADM $4$-momentum. Then, examples where the ADM formalism fails to give consist results, but the teleparallel approach does not, are given. The advantages of the teleparallel stress-energy tensor (density) over the pseudo-tensor of Landau-Lifshitz are exhibited. Finally, the difficulties in identifying the gravitational angular momentum density is discussed; it is  shown that the spatial part of the proposed angular momentum density $M^{ab}$ vanishes when the teleparallel frame satisfies the time gauge condition.
\end{abstract}


\maketitle

\newpage 

\section{Introduction}
All of the fields responsible for the four fundamental interactions of nature are supposed to carry energy and momentum, and one expects to be able to somehow quantify these energy-momenta. One of these fields is the gravitational one, which is accessible to our everyday experience.  However, the description of the gravitational energy is still a controversial subject, and most physicists believe that it is not possible to have a well-defined gravitational energy-momentum tensor density owing to the principle of equivalence \cite{Gravitation}. 

Throughout decades, many prominent scientists have tackled the task of solving this problem \cite{Einstein:1915by,10.2307/20488488,PhysRev.89.400,MOLLER1961118,Arnowitt2008,Weinberg:1972kfs,Landaufourthv2,PhysRevD.47.1407,PhysRevLett.83.1897,aldrovandi2012teleparallel,ANDP:ANDP201200272}. So far the only consensus is that 
the energy-momentum defined as surfaces terms are acceptable and have some physical meaning \cite{PhysRevLett.83.1897}. But no consensus concerning the localization of the gravitational energy, i.e, the existence of a unique and well-defined energy-momentum tensor density, has been achieved. Furthermore, even the energy-momenta defined as surface terms have many issues: there are too many giving inconsistent results and with undesirable restrictions.

An example of a $4$-momentum for the spacetime that is too limited is that due to Arnowitt-Deser-Misner (ADM) \cite{Arnowitt2008}, which seams to appear as a natural consequence of the Hamiltonian formulation of General Relativity (RG). It is limited to a very special kind of coordinate system and to asymptotically flat boundary conditions. For example, one cannot use it to evaluate the total energy of the Universe or the energy of a static and spherically symmetric spacetime in Kruskal coordinates.

On the other hand, the Hamiltonian formulation of the Teleparallel Equivalent of General Relativity (TEGR) \cite{doi:10.1063/1.530774,Maluf1999,PhysRevD.64.084014,PhysRevD.65.124001,PhysRevD.82.124035} yields an energy-momentum that does not possess those limitations. It has successfully predicted the spacetime energy-momentum in both Kruskal and Novikov coordinate systems \cite{Gonalves2021}, and also gives consistent results for the Friedmann-Lema\^itre-Robertson-Walker spacetime \cite{doi:10.1142/S021773232150125X}. In principle, its only restriction is the use of a frame that is free from artificial properties. (For more details on this later issue, see p. 20 - 54 of Ref.~\cite{formiga2022meaning}.) Nevertheless, this is not a restriction on the spacetime.

The teleparallel approach naturally gives a well-defined stress-energy tensor (density) that, in some sense, is
on the same footing as the acceleration tensor: they both depend on the tetrad field, but do not depend on the coordinate system. This tensor is probably the most promising tensor to solve the localization problem, or at least give a better understanding of it.

Because of the importance of both ADM and teleparallel approaches, it is interesting to know whether or not they are compatible. More precisely, it is important to know whether their energy-momenta contradict one another. Our purpose in this paper is to show the advantages of the teleparallel approach. In doing so, we prove that the ADM energy-momentum is a particular case of the teleparallel one, section \ref{21012022b}, and show the advantages of the teleparallel stress-energy tensor  over the Landau-Lifshitz one (section \ref{21012022c}). In addition, we prove that the spatial part of the so-called angular momentum density that arises from the Hamiltonian formulation of TEGR vanishes when the ``rigid'' frame satisfies Schwinger's time gauge (section \ref{21012022d}); we also discuss the difficulties with the interpretation of this quantity. In the last section we present the conclusions and some final remarks.

It is worth noting that in this paper we deal only with the TEGR.
In practice, the TEGR is GR written in terms of the tetrad field. The only role played by the concept of teleparallelism here is to motive the way in which Einstein's field equations are written and identify the gravitational energy. Therefore, the reader does not have to have any background on teleparallelism in order to follow this paper. (All that is necessary to understand this paper will be given in the next section.)  For those interested in modified teleparallel theories of gravity, see Refs.~\cite{Cai_2016,PhysRevD.92.104042}.

\section{The ADM and the TEGR formalisms}\label{21012022a}
In this section we present the basics features of both the ADM formalism and the TEGR theory. We start by establishing the notation and conventions.

We use the spacetime signature $(-,+,+,+)$. Greek letters represent spacetime indices, running over the values 0,1,2,3. Latin letters in the beginning of the alphabet represent tangent space indices and also run over from 0 to 4, but we use round brackets around these numbers to distinguish them from the spacetime indices  ($A_{(0)}$, for example). On the other hand, Latin indices in the middle of the alphabet run over 1,2,3 and can be used either as a spacetime index, in which case there is no bracket ($A_i$, for example), or a tangent space one ($A_{(i)}$, for example).  The four-dimensional metric and frame will be denote by $g_{\mu\nu}$ and $\e{_a}$, respectively; the three-dimensional versions will be denoted by $\tri{g}{_i_j}$ and $\tri{e}{_a}$.

\subsection{ADM formalism}
In the ADM formalism, the metric components are written as \cite{Arnowitt2008}
\begin{align}
\tri{g}{_i_j}\equiv g_{ij},\ N\equiv\left(-g^{00} \right)^{-1/2},\ N_i\equiv g_{0i},
\nonumber\\
 g_{00}=-\left(N^2-N_iN^i \right),\ N^i\equiv \tri{g}{^i^j}N_j, 
\nonumber\\ 
\tri{g}{^i^j}\ \textrm{is the inverse of}\ g_{ij},  \label{20082021a}
\end{align}
where the functions $N$ and $N_i$ are known as the {\it lapse} and {\it shift} functions. (Note that the three-dimensional $\tri{g}{_i_j}$ equals the spatial part of the four-dimensional $g_{\mu\nu}$. However, as will be clear below, $\tri{g}{^i^j}$ does not necessarily equal the spatial part of $g^{\mu\nu}$.)

The inverse components of the metric are related to the lapse and shift functions by
\begin{align}
  g^{0i}=\frac{N^i}{N^2},\ g^{00}=-\frac{1}{N^2},\   g^{ij}=\tri{g}{^i^j}-\left(\frac{N^iN^j}{N^2}\right),
\nonumber\\
\sqrt{-g}=N\sqrt{\tri{g}{}}. \label{20082021b}
\end{align}
In quantities with the prefix $^3$, the  upper spacetime indices have been raised by $\tri{g}{^i^j}$.

The ADM energy is given by the following surface integral \cite{Arnowitt2008} 
\begin{equation}
P^0=k\oint_S dS_i \left(g_{ij,j}-g_{jj,i} \right), \label{20082021c}
\end{equation}
where summation on repeated indices is assumed, and the two-dimensional surface must be at spatial infinity. 

In turn, the total field momentum is
\begin{equation}
P^i=-2k\oint_S dS_j \tensor[^3]{\pi}{^i^j}, \label{20082021d}
\end{equation}
where $\tensor[^3]{\pi}{^i^j}$ is the momenta conjugate to the $g_{ij}$; it can be written in terms of the extrinsic curvature of the hypersurface $t=$constant as (see, e.g., p. 2006 of Ref.~\cite{Arnowitt2008})
\begin{equation}
^3\pi^{ij}=-\sqrt{^3g}\left(\K{^i^j}-   \tri{g}{^i^j}\K{} \right). \label{20082021e}
\end{equation}
The extrinsic curvature $K_{ij}$ can be obtained from the spatial part of 
\begin{equation}
K_{\mu\nu}=-\nabla_{(\nu} n_{\mu)}, \label{20082021f}
\end{equation}
where $\nabla_{\nu} n_{\mu}$ is the covariant derivative, with respect to the Levi-Civita connection, of the normal vector
\begin{equation}
n_{\mu}=-N\delta^0_\mu.
\label{24082021a}
\end{equation}
(Keep in mind that, by definition, $\K{^i^j}\equiv \tri{g}{^i^p}\tri{g}{^j^q}K_{pq}$.)

The $4$-momentum defined by Eqs.~(\ref{20082021c}) and (\ref{20082021d}) are well defined only for asymptotically flat spacetimes  and coordinate systems that are asymptotically rectangular.

For more details on the ADM formalism, see Ref.~\cite{Arnowitt2008} or section 21.7 of Ref.~\cite{Gravitation}. 

\subsection{The TEGR approach}
In this section we give a brief overview of the TEGR, starting with a historical background and ending with the main quantities of interest. 

\subsubsection{Historical background}
The history of teleparallel theories dates back to Einstein's attempt to unify gravity with electromagnetism, a period that started in 1928 and ended in 1931 \cite{SAUER2006399}. Although the idea about distant parallelism was already present in the works of Weitzenb\"{o}ck, Eisenhart, and Cartan \cite{RolandW1923,Cartan1930}, it was Einstein who first applied it to a physical theory. Einstein was inspired by the sixteen degrees of freedom that the tetrad field has, six more than the metric tensor. So, he thought he could use this extra degrees of freedom to account for the electromagnetic field. However, he gave up this idea because he consider the theory to be problematic. For him,  there was too much freedom in the choice of the field equations, and it was not possible to find a tensor-like representation of the electromagnetic field (for more details, see Sauer \cite{SAUER2006399}).

Teleparallelism was revived in 1961 by M\o ller in a completely different context \cite{MOLLER1961118,MOLLER1961Mat}. M\o ller realized that the tetrad formulation of General Relativity that naturally appears in teleparallelism could help solve the problem of the gravitational energy. However, M\o ller approach did not solve the problem, because there were an infinite number of different ways of defining the energy distribution, and he did not have a fundamental way to justify choosing one over another.

After M\o ller, teleparallelism was revived  again in the 1970s by Y. M. Cho and K. Hayashi \cite{PhysRevD.14.2521,HAYASHI1977441}. They obtained teleparallelism as a gauge theory of translation. K. Hayashi showed that this gauge theory is, in fact, a theory based on the Weitzenb\"{o}ck space. 

Despite all of these efforts, the Hamiltonian formulation of the theory was initiated only in the 1990s, with the works of Maluf and collaborators \cite{doi:10.1063/1.530774,Maluf1999,PhysRevD.64.084014,PhysRevD.65.124001,PhysRevD.82.124035}. In this formulation, one is naturally led to the momentum canonically conjugated to the tetrad field and to a $4$-momentum for the spacetime. One is also led to write Einstein's field equations in a very particular way that fits the view of the Hamiltonian formulation. The common feature of Maluf's, M\o ller's and Cho's approaches is the field equations:  they worked with Einstein's field equations, i.e., they worked  with a version of teleparallelism that is equivalent to General Relativity (TEGR). However, Maluf and M\o ller wrote their field equations in a different form (their 4-momentum are also different), which means that they ended up with a different interpretation for the gravitational energy. Since Maluf's approach is based on the Hamiltonian formalism, we consider it to be the most promising one. In other words, we consider not only the $4$-momentum motivated by the Hamiltonian formalism to be the best approach to study the gravitational energy, but also the form in which Einstein's field equations are written when adapted to this view.

In the next section we write Einstein's field equations in this particular form and present the basic notions that will be used throughout this paper.

\subsubsection{Field equations and energy-momentum}
In the TEGR, Einstein's field equations are written in the form \cite{ANDP:ANDP201200272}
\begin{equation}
\pd{_\nu}\left(e\potential{^a^\lambda^\nu} \right)=\frac{e}{4k}\left(\energy{^\lambda^a}+T^{\lambda a} \right), \label{13082020a}
\end{equation}
where $k=1/(16\pi)$ in natural units, $T^{\lambda a}$ is the matter stress-energy tensor, $\energy{^\lambda^a}$ is interpreted as the gravitational stress-energy tensor, and $e=\det(\e{^a_\mu})$ is the determinant of the tetrad field $\e{^a_\mu}$. (The quantity $\e{^a_\mu}$ represents the components of the coframe in the coordinate basis, while $\e{_a^\mu}$ are the components of the frame, that is,  $\teta{^a}=\e{^a_\mu}dx^\mu$ and $\e{_a}=\e{_a^\mu}\partial_\mu$.)

The quantity $\potential{^a^\lambda^\nu}$, which transforms as a second-rank tensor field under coordinate transformations, is called superpotential and can be written as (see, e.g., section\footnote{For the original form, see Eq.~(24) of Maluf \cite{ANDP:ANDP201200272}.} 3.5.1 of Ref.~\cite{formiga2022meaning})
\begin{equation}
\potential{_a_b_c}=\frac{1}{2}\sconnection{_c_a_b}+\sconnection{^d_d_{[c}}\tensor{\eta}{_{b]a}},\label{20052019f}
\end{equation}
where 
\begin{eqnarray}
\sconnection{^a_b_c}=\frac{1}{2}\left( \torsion{_b_c^a}+\torsion{_c_b^a}-\torsion{^a_b_c} \right) \label{10092020a}
\end{eqnarray}
is the Levi-Civita spin connection; it is nothing but the Levi-Civita connection coefficients in the tetrad basis, which we have defined as  $\sconnection{^a_b_c}\equiv \bracket{\teta{^a},\nablab_b e_c }$, where $\nablab$ is the Levi-Civita connection per se. In a coordinate basis one may use $\connection{^\lambda_\mu_\nu}\equiv \bracket{dx^{\lambda},\nablab_{\mu}\partial_{\nu}}$, which are the connection coefficients of $\nablab$ in the coordinate basis $\{\partial_{\nu}\}$; the $\connection{^\lambda_\mu_\nu}$ are the well-known Christoffel symbols.

It is worth noting here that we are using the following convention. Given an object $A^\mu$, not necessarily a tensor, we define the quantity $A^a$ to be $A^a\equiv \e{^a_\mu}A^\mu$. For example, the $\torsion{^a_b_c}$ in Eq.~(\ref{10092020a}) is defined as $\torsion{^a_b_c}\equiv \e{_b^\mu}\e{_c^\nu}\torsion{^a_\mu_\nu}$, where
\begin{equation}
\torsion{^a_\mu_\nu}=\pd{_\mu}\e{^a_\nu}-\pd{_\nu}\e{^a_\mu} \label{06122021a}
\end{equation}
is basically the object of anholonomity, which is sometimes called ``the structure functions of the frame'' \cite{PhysRevD.91.084026}, or commutation coefficients of the basis $\{\e{_a}\}$ \cite{Gravitation} (be aware of possible sign differences). 

In view of the teleparallel formalism, the object of anholonomity coincides with the so-called Weitzen\"{o}ck's torsion. This is so because the Weitzen\"{o}ck connection coefficients are assumed to vanish for a particular tetrad field. In other words, there is a frame that is parallel transported everywhere via this connection. We have been calling this frame the {\it teleparallel frame}.

In the Hamiltonian formulation of the TEGR (see Maluf~\cite{ANDP:ANDP201200272} for more details), the momentum canonically conjugated to $\e{_a_\mu}$ is given by $\Pi^{a\mu}=-4ke\potential{^a^0^\mu}$. Thus, the left-hand side of Eq.~(\ref{13082020a}) is essentially the total divergence of $\Pi^{a\mu}$: taking $\lambda=0$ and using the fact that $\potential{^a^0^0}=0$, we can recast Eq.~(\ref{13082020a}) as 
\begin{equation}
-\partial_i\Pi^{ai}=e\energy{^0^a}+eT^{0a}. \label{16052022a}
\end{equation}
This justifies writing Einstein's field equations in the form given by Eq.~(\ref{13082020a}).

Integrating Eq.~(\ref{16052022a}) over the hypersurface $t=$constant, one obtains
\begin{equation}
P^a=P^a_g+P^a_M,
\label{22012022a}
\end{equation}
where we interpret $P^a\equiv -\int_Vd^3x\partial_j\Pi^{aj}$ as the spacetime energy-momentum, $P^a_M\equiv\int_V d^3x eT^{0a}$ as the matter energy-momentum, and $P^a_g\equiv\int_V d^3x e\energy{^0^a}$ as the gravitational energy-momentum, all of them defined inside the region $V$.

If we assume that there is no singularity in the region V, or, equivalently, if the singularity does not contribute to the total energy, then we can use Stokes' theorem to rewrite $P^a$ in the form
\begin{equation}
P^a=4k\oint_S dS_j e\potential{^a^0^j}. \label{20082021g}
\end{equation}
The question whether a spacetime singularity gives any contribution to the total energy is still an open problem. In fact, little effort has been made to answer this question, despite being an important one. In Ref.~\cite{Gonalves2021}, we proved that the Schwarzschild black hole singularity does not contribute to Eq.~(\ref{20082021g}). This result can be inferred from the second term in Eq.~(9) of this reference [or, equivalently, in Eq.~(22) there]: since this term corresponds to the inner boundary, the fact that it goes to zero as we approach the singularity means that the singularity does not given any contribution to the total energy, i.e., to Eq.~(\ref{20082021g}).

The $4$-momentum (\ref{20082021g}) is invariant under coordinate transformations of the three-dimensional space, i.e., coordinate transformations that do not change the time coordinate. This is a fundamental property for any $4$-momentum, because these kind of transformations do not change the state of motion of the test particles that are the constituent of the frame. ($P^a$ can, however, depend on other types of coordinate transformations\footnote{As pointed out by J. D. Norton \cite{Norton_1993}, p. 837,  for each coordinate system, there is a frame of reference whose curves coincide with the curves of constant spatial coordinates. (One says that these coordinates are adapted to this frame.) Furthermore, a frame of reference is a space filling system. We, therefore, conclude that a change of coordinates that changes the time coordinate may change the state of motion of the particles of the frame, which may alter the energy-momentum of the field.}.) The $4$-momentum (\ref{20082021g}) is also invariant under time reparametrizations and global SO(3,1) transformations. It also has the advantage of not being limited to asymptotic regions.

An important difference between the TEGR $4$-momentum and that of the ADM formalism is the role played by the tetrad field in the former. Equation (\ref{20082021g}) depends on the tetrad. Unfortunately, its dependency goes beyond the dynamics of the frame; it somehow mimics the coordinate system dependency of the ADM expression. This means that it is sensible to artificial properties of the tetrad field, i.e., properties that are not related to the state of motion of the physical system. A possible solution to this problem is to lock the tetrad axes to a physical system in a consistent way. For example, the vector field $\e{_{(0)}}$ can be locked to the timelike geodesic of freely falling particles, while the triad $\e{_{(j)}}$ can be locked to the directions of the angular momenta of three gyroscopes. (See p. 20-54 of Ref.~\cite{formiga2022meaning} for a discussion of the possible solutions to this problem.)

In order to compare Eqs.~(\ref{20082021c}) and (\ref{20082021d}) with Eq.~(\ref{20082021g}), we need to write the tetrad field in terms of the lapse and shift functions. A possible $3+1$ decomposition for the tetrad fields in the same coordinate system as that of Eq.~(\ref{20082021a}) is given by  
\begin{align}
\e{^a_i}=\tri{e}{^a_i},\ \e{^a_0}=N\eta^a+\tri{e}{^a_i}N^i,\ \eta^a=-N\e{^a^0},
\nonumber\\
\e{^a^i}=\tri{e}{^a^i}+\frac{N^i}{N}\eta^a. \label{20082021h}
\end{align}

Next, we restrict the tetrad field to Schwinger's time gauge and show that the vector field ${\bf n}\equiv n^\mu\pd{_\mu}$, where $n^\mu\equiv g^{\mu\nu} n_\nu$ and $n_\nu$ is given by Eq.~(\ref{24082021a}), coincides with $\e{_{(0)}}$ in this gauge.

\subsection{Time Gauge}\label{18012022a}
The time gauge can be characterized by demanding that $\e{_{(i)}^0}=0$. In this gauge, the following properties hold  \cite{PhysRev.130.1253}:
\begin{align}
\etime{_{(i)}^0}=0,\ \etime{^{(0)}_i}=0,\  \etime{^{(0)}_0}=\frac{1}{\etime{_{(0)}^0}}, \label{20082021i}
\\
\etime{^{(k)}_0}=-\etime{^{(0)}_0}\etime{^{(k)}_l}\etime{_{(0)}^l},\ e=\etime{^{(0)}_0}\tri{e}{}, \label{20082021j}
\end{align}
where, from now on, we will use $\etime{_a^\mu}$ to represent the components of a tetrad field that satisfies the time gauge; we have also defined $\tri{e}{}\equiv \det(\etime{^{(k)}_l})$. (Note that we do not need to change $e$ because it is independent of the tetrad, up to a sign.)

We also have the orthonormality properties
\begin{align}
\etime{_{(k)}^i}\etime{^{(k)}_j}=\delta^i_j,\ \etime{_{(i)}^k}\etime{^{(j)}_k}=\delta^j_i. \label{20082021l}
\end{align}

Applying the time gauge to a tetrad field written in the form given by Eq.~(\ref{20082021h}) yields the expressions
\begin{align}
\etime{_{(0)}^0}=1/N,\ \etime{_{(0)}^i}=-N^i/N,\ \etime{_{(i)}^0}=0,
\nonumber\\
 \etime{_{(i)}^j}=\etime[^3]{_{(i)}^j},\ e=N\,\tri{e}{}
\label{20082021m} 
\end{align}
and the coframe
\begin{align}
\etime{^{(0)}_0}=N,\ \etime{^{(0)}_i}=0,\ \etime{^{(i)}_0}=\etime{^{(i)}_j}N^j,\ \etime{^a_i}=\etime[^3]{^a_i}.
\label{20082021n} 
\end{align}

Comparing $\etime{_{(0)}^\mu}$ given by Eq.~(\ref{20082021m}) with (21.71) of Ref.~\cite{Gravitation}, we find that ${\bf n}=\etime{_{(0)}}$, where ${\bf n}=n^\mu\pd{_\mu}$ and $n^\mu=g^{\mu\nu} n_\nu$. In other words, when we assume the time gauge, the vector $\e{_{(0)}}$ becomes normal to the hypersurface of simultaneity $t=$ constant, where $t$ here is the coordinate time, and coincide with the normal vector used in the ADM formalism.

Note that impose the time gauge in the general frame (\ref{20082021h}) is not a restriction on the spacetime, but rather on the tetrad field. Therefore, we can always assume (\ref{20082021m}) and (\ref{20082021n}).

\section{The energy-momentum of the TEGR in the time gauge} \label{21012022b}
In this section we show that the $4$-momentum $P^a$  of the TEGR generalizes that of the ADM formalism when the teleparallel frame satisfies the time gauge. First, we prove that the $3$-momenta are the same for the cases in which the ADM energy-momentum holds. Then, we use the results obtained in Refs.~\cite{ANDP:ANDP201200272,Gonalves2021} to conclude that Eq.~(\ref{20082021g}) generalizes Eqs.~(\ref{20082021c}) and (\ref{20082021d}).

To find the relation between the extrinsic curvature of the hypersurface $t$ constant, where $t$ is the same time coordinate used in Eq.~(\ref{20082021a}), and the superpotential, we 
write Eq.~(\ref{20052019f}) in the form $\potential{^a^\mu^\nu}$ and take $\mu=0$. This gives  
\begin{align}
\potential{^a^0^i}=\frac{1}{2}\e{_c^i}\e{_b^0}\sconnection{^c^a^b}+\frac{1}{2}\e{^a^0}\e{_c^i}\sconnection{^b_b^c}-\frac{1}{2}\e{^a^i}\e{_b^0}\sconnection{^d_d^b}.
\label{23082021a}
\end{align}

Let us restrict Eq.~(\ref{23082021a}) to the time gauge for the case where $a=(j)$. First, we notice that the second term of Eq.~(\ref{23082021a}) vanishes for $a=(j)$, because $\etime{^{(j)}^0}=0$. Thus, we only need to focus on the other two.

From the first and the third equalities  in Eq.~(\ref{20082021m}), we see that $\etime{_b^0}\sconnection{_{\tau\!\!}^c^a^b}=\left(-1/N\right)\sconnection{_{\tau\!\!}^c^a_{(0)}}$ and, of course, $\etime{_b^0}\sconnection{_{\tau\!\!}^d_d^b}=\left(-1/N\right)\sconnection{_{\tau\!\!}^d_d_{(0)}}$; the label ``$\tau$'' indicates that $\sconnection{^a_b_c}$ has been evaluated in a frame that satisfies the time gauge. Since $\sconnection{_a_b_c}=-\sconnection{_c_b_a}$, we must have $\sconnection{_{(0)}_b_{(0)}}=0$. So, the last two expressions in the time gauge can be rewritten in the form
\begin{equation}
\etime{_c^i}\etime{_b^0}\sconnection{_{\tau\!\!}^c^a^b}=-\frac{1}{N}\etime{_{(k)}^i}\sconnection{_{\tau\!\!}^{(k)}^a_{(0)}},
\label{23082021b}
\end{equation}
and
\begin{equation}
\etime{_b^0}\sconnection{_{\tau\!\!}^d_d^b}=-\frac{1}{N}\sconnection{_{\tau\!\!}^{(k)}_{(k)}_{(0)}},
\label{23082021c}
\end{equation}
where we have contracted the first one with $\etime{_c^i}$. 

In turn, from the definition of $\sconnection{^c_a_b}$ we have\footnote{The symbol $\nabla_\mu\e{_b^\lambda}$ represents the components of the covariant derivative of $\e{_b}$ with respect to the Levi-Civita connection, $\nablab$. It is given by $\nabla_\mu\e{_b^\lambda}=\partial_\mu\e{_b^\lambda}+\connection{^\lambda_\mu_\nu}\e{_b^\nu}$.} $\sconnection{^c_a_b}=\e{^c_\lambda}\e{_a^\mu}\nabla_\mu\e{_b^\lambda}$, which can be recast as $\sconnection{^c_a_b}=\e{^c^\lambda}\e{_a^\mu}\nabla_\mu\e{_b_\lambda}$.  From the last identity, we see that $\sconnection{^{(k)}_a_{(0)}}=\e{^{(k)}^\lambda}\e{_a^\mu}\nabla_\mu\e{_{(0)}_\lambda}$. Now, from the equality\footnote{See the proof at the end of section \ref{18012022a}.}    $\etime{_{(0)}_\lambda}=n_\lambda$ and Eq.~(\ref{24082021a}), we find that $\sconnection{_{\tau\!\!}^{(k)}_{(j)}_{(0)}}=N \connection{^0_p_q}\etime{^{(k)}^q}\etime{_{(j)}^p} $, where $\connection{^\nu_\mu_\lambda}$ are the Christoffel symbols of $g_{\mu\nu}$ (the four-dimensional metric) and we have used the fact that $\etime{^{(k)}^0}=0$. On the other hand, from Eqs.~(\ref{20082021f}) and (\ref{24082021a}), we see that the extrinsic curvature takes on the form $K_{ij}=-N\connection{^0_i_j}$. Hence we have 
\begin{equation}
\sconnection{_{\tau\!\!}^{(k)}_{(j)}_{(0)}}=-K_{pq}\etime{_{(j)}^p}\etime{^{(k)}^q}.
\label{24082021b}
\end{equation}

Since $\etime{_{(i)}^j}=\tri{\tau}{_{(i)}^j}$, we can use the first equation in (\ref{20082021l}) to obtain the relation
\begin{equation}
\etime{^{(k)}^q}\etime{_{(k)}^p}=\tri{g}{^p^q}.
\label{24082021c}
\end{equation} 
Thus, contracting $k$ with $j$ in Eq.~(\ref{24082021b}) gives
\begin{equation}
\sconnection{_{\tau\!\!}^{(k)}_{(k)}_{(0)}}=-\tri{K}{},
\label{24082021d}
\end{equation}
where $\tri{K}{}\equiv K_{pq}\tri{g}{^p^q}$. 

From Eqs.~(\ref{24082021b}) and (\ref{24082021d}), we find that Eqs.~(\ref{23082021b}) and (\ref{23082021c}) with $a=(j)$  can be rewritten as $\etime{_c^i}\etime{_b^0}\sconnection{_{\tau\!\!}^c^{(j)}^b}=\left(1/N\right)\tri{K}{^{(j)}^i}$ and $\etime{_b^0}\sconnection{_{\tau\!\!}^d_d^b}=\tri{K}{}/N$, where we have used Eq.~(\ref{24082021c}) and  $\tri{K}{^{(j)}^i}\equiv K_{pq}\tri{\tau}{^{(j)}^p}\,\tri{g}{^i^q}$ (remember that $\etime{_{(i)}^j}=\etime[^3]{_{(i)}^j}$). Substituting these expressions into Eq.~(\ref{23082021a}) with $a=(j)$ and recalling that the second term vanishes for $a=(j)$, we arrive at
\begin{equation}
\potential{_{\tau\!\!}^{(j)}^0^i}=\frac{1}{2N}\left(\tri{K}{^{(j)}^i}-\tri{\tau}{^{(j)}^i}\,\tri{K}{} \right).
\label{24082021e}
\end{equation}

We know that $e=\sqrt{-g}$ (tetrads with a positive determinant), where $g$ is the metric determinant. In turn, we also know that $\sqrt{-g}=N\sqrt{\tri{g}{}}$. Hence, we have $e=N\sqrt{\tri{g}{}}$. From this identity and Eq.~(\ref{24082021e}), we obtain
\begin{align}
e\potential{_{\tau\!\!}^{(j)}^0^i}=\frac{1}{2}\sqrt{\tri{g}{}}\left(\tri{K}{^{(j)}^i}-\etime{^{(j)}^i}\,\tri{K}{} \right),
\label{25082021a}
\end{align}
where we have removed the number $3$ from the left side of the tetrad field because $\etime{_{(j)}^i}=\tri{\tau}{_{(j)}^i}$. Comparing Eq.~(\ref{25082021a})  with (\ref{20082021e}) and using the fact that $\etime{^{(j)}^p}=\tri{g}{^p^q}\etime{^{(j)}_q}$, we find that $e\potential{_{\tau\!\!}^{(i)}^0^j}=(-1/2)\tri{\pi}{^{(i)}^j}$, where $\tri{\pi}{^{(i)}^j}\equiv\etime{^{(i)}_k}\tri{\pi}{^k^j}$. Finally, using this result in Eq.~(\ref{20082021g}), we arrive at 
\begin{equation}
P_{\tau}^{(i)}=-2k\oint_S dS_j \tri{\pi}{^{(i)}^j}.
\label{25082021b}
\end{equation}

It is clear in the above expression that the teleparallel momentum in the time gauge and the ADM momentum, Eq.~(\ref{20082021d}), will coincide whenever $\tri{\pi}{^{(i)}^j}$ and $\tri{\pi}{^i^j}$ are equivalent on the two surface $S$. This equivalence happens either when $\etime{^{(i)}_k}=\delta^i_k$ or when $\tri{\pi}{^i^j}$ vanishes. (Note, however, that the ADM $P^0$ may be different from the TEGR $P^{(0)}$ even in those cases.)

Equations (\ref{20082021c}) and (\ref{20082021d}) have been defined in asymptotically flat regions and in a coordinate system where the metric components tend to those of Minkowski. In this situation, the shift functions go to zero and the lapse function goes to $1$. If we assume that the components of the teleparallel frame are written in the coordinate basis of this special coordinate system, we will certainly have $\etime{^{(i)}_k}=\delta^i_k$, ensuring the equivalence between the ADM $P^i$ and the TEGR $P^{(i)}$. A similar argument was used by Maluf et al. \cite{PhysRevD.65.124001} to show that the energy of the TEGR coincide with the ADM energy in this situation, i.e., $P^0=P^{(0)}$. Therefore, in all cases where the ADM energy-momentum is applicable, the TEGR energy-momentum can be made to coincide with the ADM version by taking a teleparallel frame that satisfies the time gauge condition asymptotically. 

There are, however, cases where the ADM $4$-momentum cannot be applied (and if applied, it fails to predict the right energy), but the TEGR $4$-momentum can and give the right answer. An example of this situation is the Kruskal spacetime \cite{Gonalves2021}: one can predict the right spacetime energy in Kruskal (also in Novikov) coordinates by using Eq.~(\ref{20082021g}), but not Eq.~(\ref{20082021c}). Furthermore, the TEGR energy-momentum is invariant under general coordinate transformations of the three-dimensional space, a property that the ADM $4$-momentum does not have.    This shows that the TEGR approach generalizes that of the ADM formalism.

\section{The TEGR energy-momentum tensor density versus the Landau-Lifshitz pseudo-tensor }\label{21012022c}
Since the stress energy pseudotensor of Landau and Lifshitz is compatible with the ADM energy and momentum \cite{Arnowitt2008}, in this section we compare it with $\energy{_\mu_\nu}$ and show the advantages of the latter.

We can use Eckart's decomposition \cite{PhysRev.58.919} for the stress-energy tensor to decompose the symmetric part of $\energy{^\mu^\nu}$. This procedure leads to the energy density $\rho\equiv \energy{_{(\mu\nu)}}\e{_{(0)}^\mu}\e{_{(0)}^\nu}=\energy{_{(0)(0)}}$ and the isotropic pressure $p\equiv(1/3)\energy{_{(\mu\nu)}}h^{\mu\nu}=(1/3)\energy{_{(0)(0)}}$, where $h_{\mu\nu}\equiv g_{\mu\nu}+\e{_{(0)\mu}}\e{_{(0)\nu}}$. (We have used the fact that $\energy{_\mu_\nu}$ is traceless.) Therefore, the TEGR predicts that the gravitational field satisfies a radiation-like equation of state, $p=\rho/3$, which is compatible with the fact that the  graviton is a massless particle. 

The above result is a consequence of the fact that any traceless stress-energy tensor satisfies this type of equation, as is obvious from the above proof. As we shall see shortly, the Landau and Lifshitz pseudo-tensor does not satisfy this property and, therefore, cannot account for the energy density of a massless field, such as the gravitational one. 

 In Landau and Lifshitz approach to Einstein's field equations, Eq.~(\ref{13082020a}) is written in the form (see, e.g., p. 138 of Ref.~\cite{bambi2018} or section 20.3 of Ref.~\cite{Gravitation})
\begin{equation}
\partial_\lambda \tau^{\mu\nu\lambda}=-g\left(l^{\mu\nu}+T^{\mu\nu} \right), \label{08122021a}
\end{equation}
where
\begin{equation}
\tau^{\mu\nu\lambda}=k\partial_\rho\left[(-g)\left(g^{\mu\nu}g^{\rho\lambda}-g^{\mu\rho}g^{\nu\lambda} \right) \right],
\end{equation}
 $l_{\mu\nu}$ is the stress-energy pseudotensor (density) of Landau-Lifshitz, and the integral of the left-hand side of Eq.~(\ref{08122021a}) yields an energy-momentum that is compatible with that of the ADM formalism. 

For the sake of simplicity and with no loss in generality, let us assume that $T^{\mu\nu}$ vanishes. In this case, we have $l^{\mu\nu}=-(1/g)\partial_\lambda\tau^{\mu\nu\lambda}$. It is clear in this expression that the $l^{\mu\nu}$ is not necessarily traceless and, therefore, it cannot always represent a massless field. As an example, consider Rindler's spacetime: $ds^2=-(1+a\xi)^2d\tau^2+d\xi^2+dy^2+dz^2$, where $a$ is the uniform acceleration of the observer at $\xi=0$. A straightforward calculation of $l^{\mu\nu}$ yields $l^{\mu \nu}=k\left(g^{\prime\prime}/g\right)(\delta^\mu_2\delta^\nu_2+\delta^\mu_3\delta^\nu_3)$ and $l=2k\left(g^{\prime\prime}/g\right)$, where $l\equiv g_{\mu\nu}l^{\mu\nu}$, $g=-(1+a\xi)^2$ is the metric determinant, and the prime represents $d/d\xi$. Using Eckart's decomposition, we find that $\rho=0$ and $p=l/3$, where we have used $h_{\mu\nu}= g_{\mu\nu}+\e{_{(0)\mu}}\e{_{(0)\nu}}$ with $\e{^{(0)}_\mu}=(1+a\xi)\delta^0_\mu$. Since $\rho=0$ and $p\neq 0$, $l^{\mu\nu}$ does not represent a radiation-like equation of state. Furthermore, the nonvanishing pressure seems to be meaningless in this case. (Note that the frame used here is the proper reference frame of Rindler's observer.)

Another problem with $l^{\mu\nu}$ is that it is sensitive to meaningless coordinate transformations (those that do not change the state of motion of the observers). To see this, take the Minkowski metric adapted to an inertial frame of reference but with spherical coordinates, i.e., $ds^2=-dt^2+dr^2+r^2d\theta^2+r^2\sin^2\theta d\phi^2$. It is straightforward to check that $l^{00}\neq 0$.

It is interesting to note that the nonvanishing of $l^{\mu\nu}$ in spherical coordinates is a good example of the problem pointed out by Laue (see, e.g., p. 233 of Ref.~\cite{NORTON1985203}): the misleading statement that the Christoffel symbols represent the gravitational field strengths. ($l^{\mu\nu}$ can be written in terms of the Christoffel symbols.) Furthermore, the nonvanishing of $l^{00}$ in spherical coordinates is not even a realization of the principle of equivalence, because the observers with constant values in those coordinates are inertial observers (they should not be able to emulate gravity locally). Any reasonable definition of a gravitational energy must not depend on a coordinate transformation that does not change the time coordinate (the state of motion).

On the other hand, $\energy{^\mu^\nu}$ vanishes in Minkowski spacetime for any coordinate system, as long as the tetrad field is either inertial or the proper reference frame of an arbitrarily accelerated observer (see theorem 2.3.1 of Ref.~\cite{formiga2022meaning}, p. 29). Another very interesting result is that  $\energy{^\mu^\nu}$ vanishes along the worldline of any observer in a curved spacetime (Levi-Civita curvature) if the teleparallel frame is the observer's proper reference frame (see theorem 2.3.2 of Ref.~\cite{formiga2022meaning}, p. 36) These results are not a realization of the principle of equivalence, because the vanishing of $\energy{^\mu^\nu}$ in these cases has nothing to do with the ``local equivalence'' between inertia and gravity. In fact, they are in agreement with the modern view, owing to Synge, that gravity is the curvature of the Levi-Civita connection (for a detailed discussion of this viewpoint, see section 7.3 of Norton~\cite{Norton_1993}). Therefore, no curvature means no gravitational field, which means no gravitational stress energy tensor.

The vanishing of $\energy{^\mu^\nu}$ along the observers worldline in the proper reference frame in a curved spacetime is more subtle: it is related to the kind of physical system that can reproduce the observer's proper coordinates. We will discuss this issue in the next section. Here let us just point out that, in this frame, Eq.~(\ref{13082020a}) becomes $4k\pd{_\nu}\left(e\potential{^a^\lambda^\nu} \right)\Bigr|_\gamma=eT^{\lambda a}\Bigr|_\gamma$ along the curve $\gamma$ (the observer's worldline). So, if the concept of locality is defined by this type of frame, then the TEGR predicts that the gravitational energy is nonlocal. (for a different view of locality, see Ref.~\cite{doi:10.1119/1.10744}; for a nice discussion of the concept of ``infinitesimal regions'', see section 10 of Ref.~\cite{NORTON1985203}.)

\section{Difficulties with the TEGR angular momentum}\label{21012022d} 
In the context of the Hamiltonian formulation of the TEGR, Maluf  has identified an object that behaves as an angular momentum and interpreted it as the gravitational angular momentum \cite{ANDP:ANDP201200272}. The gravitational angular momentum density defined by Maluf is given by
\begin{equation}
M^{ab}\equiv -4ke\left(\potential{^a^0^b}-\potential{^b^0^a} \right).
\label{19012022a}
\end{equation}
Here we discuss the challenge of interpreting $M^{ab}$ as the angular momentum density of gravity/spacetime.

First of all, it is not clear whether $M^{ab}$ is the spacetime angular momentum density or just the gravitational one (or something else). This is so because there is no known conservation equation involving $M^{ab}$ and the matter angular momentum density. To make things more complicated, we prove the following theorem.

\begin{theorem}
Let  $\tau_a$ represent a frame that satisfies the time gauge, i.e., $\etime{_{(i)}^0}=0$. In this frame, we have $M_\tau^{(i)(j)}=0$ regardless of the spacetime metric.
\end{theorem}

{\it Proof.} From Eqs.~(\ref{24082021e}) and (\ref{20082021l})-(\ref{20082021n}), one finds that $\potential{_\tau^{(j)0(k)}}=\potential{_\tau^{(k)0(j)}}$, where we have used the fact that $K_{\mu\nu}$ is symmetric. As a result,   $M_\tau^{(j)(k)}$ vanishes for any spacetime. Another way to prove that $M_\tau^{(j)(k)}$ vanishes  is as follows. From Eqs.~(\ref{20052019f}) and (\ref{10092020a}), one can easily show that $M^{ab}=-2ke\left(\torsion{^0^a^b}+\e{^b^0}T^a-\e{^a^0}T^b \right)$. In turn, from Eqs.~(\ref{20082021l})-(\ref{20082021n}) and (\ref{06122021a}), we see that $T_\tau^{0(j)(k)}$ and $\etime{^{(j)}^0}$ vanish, leading to $M_\tau^{(j)(k)}=0$.  

This result means that the angular momentum density $M_\tau^{(j)(k)}=0$ vanishes even in spacetimes such as Gödel's. If one interprets $M_\tau^{(j)(k)}$ as the spacetime angular-momentum density, then one might argue that the gravitational angular momentum density is canceling that of matter. However, there is no equation to confirm that. On the other hand, by interpret $M_\tau^{(j)(k)}$ as the gravitational density, we would have a spacetime where matter has a nonvanishing angular momentum but the gravitational field does not react to that momentum, which would be totally counterintuitive.

Another possibility is that, for some spacetimes, such as Gödel, the ideal frame to evaluate the gravitational energy does not satisfy the time gauge. It was speculated in Ref.~\cite{formiga2022meaning} (see the topic 6 in page 38) that the ideal frame to probe the gravitational energy properly should be given by a frame that is freely falling in the whole region of the spacetime where it is well defined; it must also be free from artificial properties. In many cases this idea does not conflict with the time gauge, but it is possible that it does when the gravitational field ``rotates'' locally. 

The argument in favor  of this type of freely falling frame is the following. A frame of reference is necessarily a physical system, not an abstract entity. We may neglect its stress-energy tensor, thus neglecting its effects on the spacetime curvature, because we assume that this system is made of test particles, but we cannot disregard  its qualitative properties when studying its dynamics and the gravitational energy. For instance, the Schwarzschild coordinates are related to an acceleration that is necessary to keep the test particles at rest in the radial coordinate; this coordinate system, and therefore the frame adapted to it, assumes the existence of an external (nongravitational) force. In other words, the stress-energy tensor of both the system that accelerates the test particles and the test particles themselves are assumed not to interfere significantly with the background geometry. However, qualitative properties such as ``the origin of the frame is accelerated'' can never be neglected. So, it is natural to think of two possibilities. 
The first one is to assume that these nongravitational interactions embedded in the frame prevent us from properly interpreting $\energy{^\mu^\nu}$ as the gravitational energy. The second, and perhaps the most desirable one, is that $\energy{^\mu^\nu}$ does represent the gravitational energy in any frame free from artificial properties, but the nongravitational forces change the gravitational energy in a way that may be hard to understand, at least when neglecting the effects of these forces on the background geometry. Therefore, the safest approach seems to be that which takes $\energy{^\mu^\nu}$ as the gravitational stress-energy tensor only when the frame is made up of freely falling particles.

To support the idea that these nongravitational interactions change the behavior of gravity, one just need to realize that the coordinate systems that is usually used to write an arbitrary metric in the Minkowskian form locally are all adapted to particles that are accelerated (under nongravitational interaction), with the exception of that at the origin. (For more details, see appendix \ref{31052022c}.) On the other hand, it seems that coordinate systems adapted to freely falling test particles do not allow the metric to take the Minkowskian form along the trajectory of an arbitrary freely falling particle if the spacetime is curved. Therefore, it is natural to assume that the best way to probe the gravitation energy is using a system of freely falling particles as the reference frame. (This does not mean that this type of frame is privileged; it is just a convenience for identifying and interpreting the gravitational energy.)

In finding a freely falling frame that is free from artificial properties, the minimal set of assumptions we have to make is
\begin{enumerate}
\item The acceleration tensor must vanish, i.e., $\sconnection{^a_{(0)}_b}=0$.
\item The Levi-Civita connection coefficients $\sconnection{^a_b_c}$ must vanish when the curvature tensor of this connection vanishes (absence of gravity).  
\end{enumerate}
We leave the analysis of $M^{ab}$ in these type of frames for a future work. 

\section{Conclusions and final remarks}\label{21012022e}

We have seen that the energy-momentum of the TEGR is more general than that of the ADM formalism. It gives the same results as the ADM one when the latter is applicable, but can go way beyond that. It predicts the right energy of both Kruskal and Novikov spacetimes, it does not need  asymptotically flat boundary conditions (see, e.g., the cosmological case in Ref.~\cite{doi:10.1142/S021773232150125X}) and can be used in finite regions of the spacetime.

Since the teleparallel stress-energy tensor is traceless, it satisfies a radiation-like equation of state. On the other hand, the Landau-Lifshitz pseudo-tensor is not traceless and cannot describe massless fields.

We have also seen that the quantity that is in general associated with the spacetime (or gravitational) angular momentum density in the context of the TEGR, $M^{ab}$, satisfies the relation $M^{(j)(k)}=0$ if the teleparallel frame satisfies the time gauge. (Note that this result is independent of the coordinate system.) This means that there is no angular momentum in this case, for whatever this angular momentum is supposed to be. 

Is $M^{(j)(k)}$ the angular momentum density of gravity or of the spacetime? Is it a residual angular momentum of the frame?   We leave a deeper analysis of this issue for a future work.

As a final remark, we would like to point out the possibility that the advantages of the $4$-momentum of the TEGR over that of the ADM approach extends to the Hamiltonian formulation as a whole. If that is the case, then the Hamiltonian formulation of the TEGR would be an improvement of the ADM formulation. 

\appendix
\section{Local inertial frame}\label{31052022c}
Denoting the Fermi normal coordinates by $x^\mu=(ct,x^j)$, we can write the metric components accurate to second order in $|x^j|$ in the form  $g_{\mu\nu}=\eta_{\mu\nu}+h_{\mu\nu}$ (see, e.g., p. 332 of Ref.~\cite{Gravitation}), where $\eta_{\mu\nu}$ is the Minkowski metric and
\begin{align}
h_{00}=-R_{0i0j}(t)x^ix^j,\quad h_{0i}=-\frac{2}{3} R_{0jik}(t)x^jx^k,
\\
h_{ij}=-\frac{1}{3} R_{ikjl}(t)x^kx^l, \label{31052022a}
\end{align}
where $R_{\alpha\beta\mu\nu}(t)$ are evaluated along the worldline $x^\mu=(t,0)$.

The coordinates $(ct,x^j)$ are the proper coordinates of the  observer at $x^j=0$. Along the observer's worldline, the metric becomes $\eta_{\mu\nu}$ and the connection coefficients vanish; hence, this proper reference frame is a local Lorentz frame (a local inertial frame of reference). Next, we show that the physical system which realizes this coordinate system and, of course, the frame adapted to it, is not a pure gravitational system.

It is not clear whether the concept of a tetrad field adapted to a certain coordinate system leads to a unique tetrad. However, we do not need a unique tetrad to prove that the curves with constant values of $x^j$ are accelerated for $x^j\neq 0$. Any frame that is adapted to the test particles with constant values of $x^j$ will be sufficient.

A frame that is adapted to $(ct,x^j)$ and satisfies the time gauge is
\begin{align}
\etime{^{(0)}_\mu}\approx (1-\frac{1}{2}h_{00} )\delta^0_\mu,\quad \etime{^{(i)}_0}\approx h_{0i},
\nonumber\\
\etime{^{(i)}_j}\approx \delta_{ij}+\frac{1}{2}h_{ij},\quad \etime{_{(0)}^0}\approx 1+\frac{1}{2}h_{00},
\nonumber\\
\etime{_{(0)}^i}\approx -h_{0i},\ \etime{_{(i)}^0}\approx 0,\ \etime{_{(i)}^j}\approx \delta_{ij}-\frac{1}{2}h_{ij}. \label{02062022a}
\end{align}
In principle, we could use $\etime{_{(0)}^\mu}$ as given by Eq.~(\ref{02062022a}) to calculate the acceleration of the particles of the frame. However, there is a simpler way of doing that: we can calculate the acceleration of the curves with constant values of $x^j$ in  the hypersurface of simultaneity that is orthogonal to $u^\mu=dx^\mu/d\tau$, rather than that that is orthogonal to $\etime{_{(0)}}$;  $\tau$ is the proper time of the particle with the worldline $x^\mu(\tau)$. One can show that the acceleration calculated with $u^\mu$ equals that calculated with  $\etime{_{(0)}}$ to first order in $|x^j|$. That is enough for our purpose.

The $4$-velocity takes the form $u^\mu=c (dt/d\tau) \delta^\mu_0$, where $dt/d\tau\approx 1+h_{00}/2$. Since the acceleration is given by $a^\mu=u^\nu\nabla_\nu u^\mu$, we have
\begin{equation}
a^\mu=c^2\frac{dt}{d\tau}\left(\frac{\partial}{\partial x^0}\frac{dt}{d\tau}\delta^\mu_0+\connection{^\mu_0_0}\frac{dt}{d\tau} \right), \label{31052022b}
\end{equation}
where $\connection{^\mu_\alpha_\beta}$ are the Christoffel symbols. From Eq.~(18.2) in Ref.~\cite{Gravitation}, we find that $\connection{^\mu_0_0}\approx \eta^{\mu\nu}\partial_0h_{0\nu}-(1/2)\eta^{\mu\nu}\partial_\nu h_{00}$. Hence, we consider the approximation $\connection{^\mu_0_0}(dt/d\tau)\approx \connection{^\mu_0_0}$. On the other hand, the first term in Eq.~(\ref{31052022b}) can be approximated by $\partial_0(dt/d\tau)\approx (1/2)\partial_0 h_{00}$. Using these expressions in Eq.~(\ref{31052022b}), one finds that $a^0\approx 0$ and $a^i\approx c^2 R_{0i0j}(t)x^j$ to first order. Therefore, the test particles which compose the frame are not free, except for the one at $x^j=0$; they are under the action of a nongravitational force.

In practice, the Fermi normal coordinates are realized by `rigid' bodies with nongravitational forces that fight the so-called tidal `forces'. To be more precise, the forces responsible for the acceleration $a^i$ prevents the particles from following geodesics. So, it seems clear that a system of coordinates whose positions are marked by freely falling particles will not always allow the metric to become that of Minkowski along a timelike geodesic.

\section*{Acknowledgments}
Victor Gon\c calves acknowledges CNPq for financial support.

%


\end{document}